\documentclass[12pt]{iopart}
\usepackage{epsfig}
\usepackage{graphics}
\usepackage{graphicx}
\usepackage{epstopdf}
\usepackage[dvipsnames]{xcolor}
\DeclareGraphicsExtensions{.jpeg,.eps}

\begin{document}

\title[Thermodynamic origin of the Contact]{The thermodynamic origin of the Contact and its relation to the gap in the BEC-BCS crossover}

\author{V\'{\i}ctor Romero-Roch\'{\i}n} 

\address{Instituto de F\'{\i}sica. Universidad Nacional Aut\'onoma de M\'exico. \\
Apartado Postal 20-364, 01000 M\'exico, D.F., Mexico}
\ead{romero@fisica.unam.mx}

\date{\today}

\begin{abstract}
As can be inferred from present experiments in ultracold gases, the scattering length is a quantity that determines the thermodynamic state of the gas. As such, there exists a conjugate thermodynamic to it. Here, we show that the recently introduced {\it contact}  is the  conjugate of the inverse of the scattering length. We find that this identification allows for a derivation of essentially all the known results regarding the contact. Using the mean-field theory for the Bose-Einstein (BEC) to Bardeen-Cooper-Schriefer (BCS) crossover, we  also find that the contact is proportional to the square of the gap.  We analyze in detail both a homogenous balanced mixture of fermions and its inhomogenous counterpart in a harmonic trap.
\end{abstract}


\maketitle

\maketitle

\section{Introduction}

The experimental realization of the crossover between a molecular Bose-Einstein condensate (BEC) and an atomic Bardeen-Cooper-Schriefer state (BCS) in fermionic alkali vapours, such as $^{6}$Li and $^{40}$K, is one of the most fundamental and far reaching results of the current ultracold quantum fluids research. The route to this crossover of macroscopic states started with the early observations of molecular BEC\cite{OHara-Sc2002,Greiner-Na2003,Jochim-Sc2003,Zwierlein-PRL2003,Bourdel-PRL2004} continuing to the observation of actual Fermi pairing in the BCS side\cite{Regal-PRL2004,Zwierlein-PRL2004,Chin-Sc2004,Partridge-PRL2005} that directly verified the crossover, and in the meantime, the superfluid nature of the quantum fluids was established\cite{Kinast-PRL2004,Bartenstein-PRL2004,Zwierlein-Na2005,Zwierlein-Sc2006}. Furthermore, a whole class of spectroscopic techniques have been developed to enquire about the elementary excitations spectra\cite{Veeravalli-PRL2008,Schunck-Na2008,Stewart-Na2008,Gaebler-NaP2010} which not only should verify the superfluid characteristics of the state but should also shed light on their true microscopic details. Of additional relevance, and the matter of the present article, is the study of thermodynamic properties\cite{Kinast-Sc2005,Partridge-Sc2006,Luo-PRL2007,Shin-PRL2008,Schirotzek-PRL2008} and, in particular, of direct measurement of the {\it contact} by Stewart et al.\cite{Stewart-PRL2010}, a thermodynamic variable we discuss below. We refer to the review by Giorgini et al.\cite{Giorgini-RMP2008} for further discussion and additional references. The BEC-BCS crossover occurs as a consequence of the change of the atomic scattering length by means of external magnetic fields, with the concomitant presence of a two-channel Feschbach resonance\cite{Kohler-RMP2006,Chin-RMP2010}, that permits a switch between the two extreme states. This phenomenon, as originally predicted by Eagles\cite{Eagles-PR69} and Leggett\cite{Leggett-80}, can also be theoretically studied considering a one-channel potential resonance that allows for the change of sign of the scattering length, indicating the existence of the molecular BEC and atomic BCS.
Among the many studies sparked by these experiments, we are concerned here with the appearance of a {\it new} variable called the {\it contact}. This was originally introduced in a series of papers by Tan\cite{Tan-AP2008-1,Tan-AP2008-2,Tan-AP2008-3}, and further analyzed by others\cite{Braaten-PRL2008,Zhang-PRA2009,Werner-EPJ2009,Kunhle-ArX2009}, and it has proved to be of relevance in the determination and understanding of BEC-BCS crossover. 

The contact, as introduced by Tan\cite{Tan-AP2008-1}, tells us about the asymptotic behavior  of the wavevector occupation number of fermions $n_k$, and it is defined as
\begin{equation}
C = \lim_{k \to \infty} k^4 n_k .\label{c1}
\end{equation}
That is, since the contact approximation of the interatomic potential $u(\vec r)$, implemented by replacing
\begin{equation}
u(\vec r) \approx \frac{4 \pi \hbar^2 a}{m} \delta(\vec r) ,\label{a1}
\end{equation}
with $a$ the scattering length, necessarily introduces difficulties as $\vec r \to 0$, Tan found that this produced a divergent but well characterized $k^4$ behavior of the occupation number as $k \to \infty$. This,  in turn, opens a door to ``correct" the well-kown ultraviolet divergences produced by the contact approximation (\ref{a1}). Tan called the limit (\ref{c1}) the {\it contact}. Further analysis\cite{Tan-AP2008-1} showed that an {\it adiabatic} change of the scattering length $a$, relates the contact to the energy $E$ of the system by means of the relationship,
\begin{equation}
C = - \frac{\hbar^2}{4 \pi m} \frac{1}{V} \left(\frac{\partial E}{\partial \eta}\right)_S, \label{c2}
\end{equation}
where $V$ is the volume of the system, and the derivative is taken at entropy $S$ constant. Here and throughout, we shall use $\eta = 1/a$ as the inverse of the scattering length.

In this article we point out a further and very simple connection between the scattering length $a$ and the contact $C$, a relationship that seems to have gone unnoticed. We shall argue that since $a$, or better $\eta = 1/a$, is a {\it bona fide} intensive thermodynamic state variable, it must have an extensive conjugate variable. We show here that this variable is the contact $C$ (times the volume $V$). We shall see that expressions (\ref{c1})  and (\ref{c2}), among other properties, follow from an analysis of general thermodynamic considerations and of this identification. To illustrate our results, we shall analyze in  detail all the thermodynamic predictions of the mean-field theory\cite{Leggett-80} of the BCS-BEC crossover at zero temperature, $T = 0$. As we shall see, the results drawn from this approximate theory are in reasonable agreement with both other more precise theories\cite{Werner-EPJ2009} and recent experiments\cite{Partridge-PRL2005,Stewart-PRL2010}. This work, we believe, adds to the comprehension and potential use of the contact variable.

\section{The thermodynamic origin of the contact}

The observation we made here is based on the fact that any mechanical variable of a system that can be externally varied, and that can change the current thermodynamic state of the system, is in itself a thermodynamic state variable. As an example, any adiabatic change of an internal parameter of the Hamiltonian of the system (keeping, say, the volume and number of particles constant) necessarily results in an increase or decrease of the temperature of the system; it is thus a thermodynamic parameter that determines the macroscopic state of the system. Let us therefore take $\eta = 1/a$ as a thermodynamic variable. Using the inverse of $a$ is a matter of convenience. By its mere nature, $\eta$ is an intensive variable, namely it does not scale with the size of the system.

Since most of the theoretical analyses of many-body systems is performed in the grand canonical ensemble, let us first consider the grand potential $\Omega$ of a {\it homogenous} gas. Such a function is a fundamental relationship that yields all the thermodynamics of the system\cite{LandauI} in terms of the following variables $\Omega = \Omega(V,T,\mu,\eta)$,
where $V$ is the volume of the system, $T$ its temperature, $\mu$ the chemical potential, and we have assumed an additional dependence on $\eta$. Being a function of $(V,T,\mu,\eta)$ this is equivalent to state that
\begin{equation}
d\Omega = - p dV - S dT - N d\mu - {\cal C} d\eta, \label{dO}
\end{equation}
which amounts to identify (or define) the entropy $S$, the pressure $p$, the (average) number of particles $N$, and an additional variable, the contact ${\cal C}$, as partial derivatives of $\Omega$. In particular, the contact is
\begin{equation}
{\cal C} = - \left(\frac{\partial \Omega}{\partial \eta}\right)_{V,T,\mu} .\label{dervs}
\end{equation}
Thus, just as  $(p,V)$, $(T,S)$,  and $(\mu,N)$, $(\eta,{\cal C})$ are a pair of conjugate variables.  $\eta$ being intensive implies that ${\cal C} = {\cal C}(V,T,\mu,\eta)$ is an extensive variable. As shown below, it is related related to Tan's definition of the contact $C$ as ${\cal C} = (\hbar^2/ 4\pi m) CV$. That is, Tan's $C$ is an intensive variable, a contact density (or contact ``intensity"), and in this case is a function  $C = C(T,\mu,\eta)$.

The relationship to statistical physics serves to find further properties of ${\cal C}$ and $\eta$. Recall that in the grand canonical ensemble, the state is given by the density matrix
\begin{equation}
\hat \rho = \frac{1}{\Xi} \exp(\frac{\mu}{k_B T}\hat N - \frac{1}{k_B T}\hat H) ,\label{rho}
\end{equation}
where $\hat N$ is the number of particles operator and $\hat H$ the Hamiltonian of the system. This Hamiltonian depends parametrically on the scattering length $a$, as we further specify below. $\Xi$ is the grand partition function,
\begin{equation}
\Xi = {\rm Tr} \exp(\frac{\mu}{k_B T}\hat N - \frac{1}{k_B T}\hat H) .\label{Xi}
\end{equation}
The grand potential is $\Omega = - kT \ln \Xi$ and the entropy is given by
\begin{equation}
S = -k_B {\rm Tr}\>  \hat \rho \ln \hat \rho.
\end{equation}
From this we obtain, by substituting (\ref{rho}) and (\ref{Xi}),
\begin{equation}
TS = E - \mu  N -  \Omega.\label{Euler}
\end{equation}
where $E = \langle \hat H \rangle$ is the average energy. The expression for $S$ allows to solve for $E = \Omega + TS - \mu N$, that with the use of (\ref{dO}) yields,
\begin{equation}
dE = TdS - p dV + \mu dN - C d\eta.
\end{equation}
From this expression one finds that the contact is also given by,
\begin{equation}
{\cal C} = - \left(\frac{\partial E}{\partial \eta}\right)_{S,V,N} ,
\end{equation}
which is the known expression (\ref{c2}) also derived by Tan\cite{Tan-AP2008-1}. It is the so-called adiabatic definition of the contact. It is clear that if one considers the Helmholtz free energy, $F = E - TS$, the contact is now ${\cal C} = - (\partial F/\partial \eta)_{N,V,T}$, and so on for any other thermodynamic potential.

We now specialize to a fermionic many-body system with two-body interactions, and assume that the collisions can occur only  by one open channel. Below we comment on the case of two-channel collisions\cite{Zhang-PRA2009,Werner-EPJ2009}. In general, any real interatomic potential $u(\vec r)$ depends at least on two parameters, say the depth of the potential $u_0$ and its range $r_0$. At low energies or temperatures, both parameters can be replaced by the scattering length $a$, and accordingly, the potential is replaced by the so-called contact potential, as given by eq.(\ref{a1}). As it is well known this approximation is plagued with ultraviolet divergences and several schemes have been developed to repair them. Incidentally, Tan's work on the contact variable is another instance to deal with this problem. As we shall return below, the replacement of the several parameters of the interatomic potential by solely the scattering length has very strong implications on the form of the thermodynamic potentials. In any case, the Hamiltonian in this approximation may be written as
\begin{equation}
\hat H = - \sum_{\sigma} \psi_\sigma^\dagger \frac{\hbar^2}{2 m}   \nabla^2 \psi_\sigma + \frac{4 \pi \hbar^2 a }{ m} \psi_1^\dagger \psi_2^\dagger \psi_2 \psi_1 ,\label{H}
\end{equation}
where we have already considered a mixture of fermions, say, in different internal states $\sigma =$ 1 and 2, $\psi_1$ and $\psi_2$ being their field operators. The gas is assumed to be confined in a rigid box of volume $V$ and it is, therefore, homogenous. Using the results above, eqs. (\ref{dervs}) and (\ref{Xi}) with Hamiltonian (\ref{H}), one immediately finds that the contact variable is,
\begin{equation}
{\cal C}  =  \frac{4 \pi \hbar^2}{m} \int \langle  a^2 \psi_1^\dagger \psi_2^\dagger \psi_2 \psi_1 \rangle d^3 r .\label{CBra}
\end{equation}
where the average can be taken in any ensemble, even though we obtained it in the grand ensemble. This result was originally derived by  Braaten and Platter\cite{Braaten-PRL2008}. These authors argue that a way of avoiding the ultraviolet divergence in (\ref{CBra}) is by replacing $a$ by $g(a) = a/(1 - 2a \Lambda/\pi)$ with $\Lambda$ a cutoff parameter. Moreover, after a careful analysis, they find that the above value of ${\cal C}$ is precisely the original Tan definition of the contact $C = {\cal C}/V$ as given in equation (\ref{c1}), namely, as the coefficient of the $k^{-4}$ tail of the number distribution.

If the gas is confined by an inhomogenous potential, as in the current experiments with ultracold gases, say in a harmonic trap\cite{aclara},
\begin{equation}
V_{ext}(\vec r) = \frac{1}{2} m \omega^2 r^2 ,
\end{equation}
one must add to the Hamiltonian (\ref{H})  a term $V_{ext}(\vec r) \sum_\sigma \psi_\sigma^\dagger  \psi_\sigma$. This addition is by no means trivial since it modifies the thermodynamic description of the system. As discussed in Refs.\cite{Seldam,Dalfovo-RMP1999,VRR-PRL2005,VRR-BJP2005,Sandoval-PRE2008}, among others, the thermodynamic limit is now $N \to \infty$ $\omega^3 \to 0$, with $N \omega^3 $= constant. This yields the ``generalized" volume ${\cal V} = 1/\omega^3$ as an extensive thermodynamic variable that replaces the usual volume $V$ of a rigid-box potential. In addition there appears a conjugate ``generalized" pressure ${\cal P}$ that replaces the hydrostatic pressure $p$. The details of these variables as well as their physical interpretation has been given at length in Refs.\cite{VRR-PRL2005,VRR-BJP2005,Sandoval-PRE2008}. Here, suffice to say that the grand potential of the trapped gas is a function  $\Omega = \Omega({\cal V},T,\mu,\eta)$. The contact is simply,
\begin{equation}
{\cal C} = -  \left(\frac{\partial \Omega}{\partial \eta}\right)_{{\cal V},T,\mu} \label{CHar}
\end{equation}
and it is now a function of $({\cal V},T,\mu,\eta)$. Similarly to the homogenous case, eq.(\ref{Euler}), one has the Euler relationship, $\Omega = E - TS - \mu N$, and the generalized pressure may be identified as $\Omega = - {\cal P V}$, namely as the grand potential per unit of generalized volume. We shall use this result below.

With the use of the local density approximation (LDA) one can calculate all thermodynamic variables in the trapped gas. In particular, following the procedure of Ref.\cite{Sandoval-PRE2008}, one can show that if one has knowledge of the contact density $c_h(T,\mu,\eta) = {\cal C}(V,T,\mu,\eta)/V$ of the homogenous case, the contact in the trapped gas can be obtained by LDA prescription,
\begin{equation}
{\cal C}({\cal V},T,\mu,\eta) = \int c_h(T,\mu-V_{ext}(\vec r),\eta) d^3 r.
\end{equation}
This result will be used in section 4.

If one considers two-channel collisions, such as those involved by a Feschbach resonance, Zhang and Leggett\cite{Zhang-PRA2009} and Werner et al.\cite{Werner-EPJ2009} have shown that the number of bosonic molecules in the closed channel is also related to the contact ${\cal C}$, up to a quantity that depends crucially on the two-body interactions involved in the resonance. Werner et al.\cite{Werner-EPJ2009} find very good agreement with the experimental results by Partridge et al.\cite{Partridge-PRL2005}. Although we shall not consider this case here, it is of interest to mention that a calculation of the contact with a single channel, as performed in the present article, gives reliable information on a quantity that also depends on the details of two-channel collisions. Below we return to this point where we discuss the ``universality" of the obtained thermodynamics.

\section{Reduced variables and ``universality"}

So far, we have just shown how all the known results concerning the contact follow directly from the identification of $\eta = 1/a$ and ${\cal C}$ as conjugate thermodynamic variables. Here, we would like to make some general considerations regarding the functional dependence of the different variables by assuming the contact approximation of the interatomic potential, eq.(\ref{a1}). That is, we assume that the interatomic interaction depends solely on the scattering length $a$.

Let us analyze first the homogenous case. In the grand canonical ensemble the independent variables are $(V,T,\mu,\eta)$ with only two parameters, Planck constant $\hbar$ and the atoms mass $m$. We note that the number of particles is a dependent variable $N = N(V,T,\mu,\eta)$. Thermodynamics ensures that one can single-value invert $\mu = \mu(n,T,\eta)$ with $n = N/V$ the homogenous particle density. Thus, any other thermodynamic variable can be written in terms of $(N,V,T,\eta)$, for instance, the contact ${\cal C} = {\cal C} (N,V,T,\eta)$ or the energy $E = E(N,V,T,\eta)$. Now, as long as the particle density is never taken as zero, $n$ is always positive. Therefore, one can use $\hbar$, $m$ and $n$ to adimensionalize or reduce all the other variables. Given the fact that we are dealing with fermions, it proves convenient to use the Fermi energy and momentum
\begin{equation}
\epsilon_F = \frac{\hbar^2 k_F^2}{2m} \>\>\>\>{\rm and}\>\>\>\>k_F = (3 \pi^2 n)^{1/3} \label{Fh},
\end{equation}
and the mass $m$, as the quantities to define dimensionless reduced variables. We shall place a ``tilde'' $\sim$ on top to denote reduced variables in the homogenous case. For instance, for intensive variables,
\begin{equation}
\tilde \mu = \frac{\mu}{\epsilon_F} \>\>\>\>\tilde T = \frac{kT}{\epsilon_F}\>\>\>\>\tilde \eta = \frac{\eta}{k_F}  .\label{redh1}
\end{equation}
For extensive variables we will take them as per particle, namely,
\begin{equation}
\tilde \omega = \frac{\Omega}{N \epsilon_F}\>\>\>\>\tilde e = \frac{E}{N\epsilon_F}\>\>\>\>\tilde c = \frac{{\cal C}k_F}{N\epsilon_F} \label{redh2}
\end{equation}
and so on. We note that the number density becomes a number,
\begin{equation}
\tilde n = \frac{n}{k_F^3} = \frac{1}{3\pi^2} .
\end{equation}

Hence, since extensive quantities are of the form $X = Nx(n,T,\eta)$ and intensive ones $\xi = \xi(n,T,\eta)$, then, it must be true that all the reduced variables depend {\it only} on $\tilde T$ and $\tilde \eta$, namely, $\tilde \mu = \tilde \mu(\tilde T,\tilde \eta)$, $\tilde e = \tilde e(\tilde T,\tilde \eta)$, $\tilde c = \tilde c(\tilde T,\tilde \eta)$, and so on. This indicates that the dependence is ``universal", as in a law of corresponding thermodynamic states\cite{LandauI}. However, we want to make this point clear: this dependence is only possible because the Hamiltonian has only two parameters, $\hbar$ and $m$ ($a$ is a variable not a parameter). If the interatomic potential had one more parameter in addition to $a$, then this ``universal" dependence will not be possible. That is, an explicit dependence on the density will still occur. Recently, Ho\cite{Ho} introduced a Universality Hypothesis in which, near the unitarity limit $|a| \to \infty$ or $\eta \to 0$, all interatomic potential parameters become irrelevant except the scattering length $a$. This implies, as we see here, that all thermodynamic variables, for {\it fixed} values of $\tilde T$ and $\tilde \eta$, behave, up to a numerical constant, as if the system were an {\it ideal} Fermi gas. Here, by assuming at the outset that the only relevant interatomic potential quantity is the scattering length, we find that such a hypothesis is always valid. In real systems, and away from unitarity, one expects a dependence, maybe weak, on other interatomic potential parameters. In this connection is important to mention the recent work by Zhang and Leggett\cite{Zhang-PRA2009}, where they argue that near unitarity, where $|a|$ is very large and becomes the dominant length, the many-body physics is dictated by the contact potential and, thus, its thermodynamic contribution is actually universal, thus lending validity to Ho's hypothesis. Zhang and Leggett, though, are very explicit in the fact that certain quantities are also influenced by the details of the two-body physics, such as the number of bound pairs or molecules in the closed channel of a Feschbach resonance. As it turns out, the number of bound pairs in the closed channel is found to be directly proportional, both, to a quantity that depends on the details of the two-body interatomic potentials {\it and} to the contact ${\cal C}$, that being a many-body thermodynamic quantity is independent of the details of the collision. Thus, two-body physics matters, but a one-channel calculation of thermodynamic properties, such as the contact, suffices to determine the many-body contributions.

For the case of a gas trapped in a harmonic potential the situation is very similar. First, we note that the number of particles is $N = N({\cal V},T,\mu,\eta)$ with ${\cal V} = 1/\omega^3$. Again, we solve for the chemical potential $\mu = \mu(N/{\cal V},T,\eta)$, and express all other thermodynamic quantities as functions of $(N,{\cal V},T,\eta)$, say ${\cal C} = {\cal C}(N,{\cal V},T,\eta)$. Similarly, we introduce the Fermi energy and momentum {\it in the trap} as,
\begin{equation}
\varepsilon_F = \hbar (3 N \omega^3)^{1/3} \>\>\>\>{\rm and}\>\>\>\>\kappa_F = \left(\frac{2m \varepsilon_F}{\hbar^2}\right)^{1/2} .\label{Finh}
\end{equation}
We see that the combination $N \omega^3 = N/{\cal V}$ is a generalized particle density for the harmonic trap. We define dimensionless reduced variables in the trap, denoted with a ``hat"  $\land$, in an analogous way to the homogeneous case,
\begin{equation}
\hat \mu = \frac{\mu}{\varepsilon_F} \>\>\>\>\hat T = \frac{kT}{\varepsilon_F}\>\>\>\>\hat \eta = \frac{\eta}{\kappa_F}  \label{redt1}
\end{equation}
for intensive variables, while for extensive ones
\begin{equation}
\hat \omega = \frac{\Omega}{N \varepsilon_F}\>\>\>\>\hat e = \frac{E}{N\varepsilon_F}\>\>\>\>\hat c = \frac{{\cal C}\kappa_F}{N\varepsilon_F},\label{redt2}
\end{equation}
and so on. Again, any ``hat" variable thus defined depends only on $\hat T$ and $\hat \eta$, i.e. $\hat f = \hat f(\hat T,\hat \eta)$.

\section{Mean-field BEC-BCS crossover at $T = 0$}

Further explicit results can now be obtained by appealing to the mean-field BCS-BEC theory\cite{Eagles-PR69,Leggett-80} at zero temperature $T = 0$, as clearly described by Leggett\cite{Leggett-80}. Within this approximation, one finds that the grand potential $\Omega$ is given by,
\begin{equation}
\Omega = \sum_k \left[ (\epsilon_k - \mu) - \sqrt{(\epsilon_k - \mu)^2 + \Delta^2} \right] + \frac{1}{2}\Delta^2 \sum_k \frac{1}{\epsilon_k} 
 - V \frac{m}{4 \pi \hbar^2} \Delta^2 \eta ,
\label{OMF}
\end{equation}
where $\epsilon_k = \hbar^2 k^2/2m$ and the ``gap" $\Delta$ is a function of $(T,\mu,\eta)$ given by the transcendental equation,
\begin{equation}
\frac{1}{V}\sum_k \left[\frac{1}{\sqrt{(\epsilon_k - \mu)^2 + \Delta^2}} - \frac{1}{\epsilon_k}\right] = -  \frac{m}{2
\pi \hbar^2}  \eta .\label{gap}
\end{equation}
These two equations yield, as expected, $\Omega = \Omega(V,\mu,\eta)$. The thermodynamic limit is assumed to be used in the form $\sum_k \to V/(2\pi)^3 \int d^3 k$.
Few words are in order here. This form of the grand potential is obtained by considering that the ground state of the two-species fermion mixture is given by the BCS variational state\cite{BCS}. And most importantly, we recall that we have already included in (\ref{OMF}), by hand,  the ``counterterms" needed  to avoid the ultraviolet divergences that the contact approximation (\ref{a1}) necessarily introduces. Strictly speaking, the mean-field approximation is only valid in the weakly-interacting regime, which requires that $N |a|^3/V \ll1$. That is, it should only be valid in the limit of very large $|\eta |$; this immediately prohibits to use the theory in the strongly interacting crossover region $\eta \approx 0$. Nevertherless, as we shall see below, one finds results that agree reasonably well with current experimental results\cite{Partridge-PRL2005,Stewart-PRL2010}. We have not considered finite temperatures here since, on the one hand our purpose is to exemply the main results, and on the other, it is well known that this approach fails in considering correctly all the thermal excitations\cite{Engelbrecht-PRB1997,Ohashi-PRL2002,Javanainen-PRL2005,Chen-PRL2005,Romans-PRL2005}. 

We are now in a position to calculate all the thermodynamic properties of the system by using $\Omega$ as given by (\ref{OMF}) and (\ref{gap}), as a function of $(V,\mu,\eta)$ and by recalling that $S$, $p$, $N$ and ${\cal C}$ are given by its partial derivatives. We are particularly interested in $N$ and ${\cal C}$. Recall that the number of fermions in each species is $N/2$. One finds,
\begin{equation}
N = \sum_k \left[ 1 - \frac{\epsilon_k - \mu}{ \sqrt{(\epsilon_k - \mu)^2 + \Delta^2} } \right]  \label{N}
\end{equation}
and
\begin{equation}
{\cal C} = V \frac{m}{4 \pi \hbar^2} \Delta^2 .\label{C}
\end{equation}
This second expression is an important result of this article, since one finds that the contact per volume $C/V$ is, apart from constant factors, the square of the gap $\Delta$. From the expression for $N$, eq.(\ref{N}), one can infer the occupation number of  fermions (of any species) in state $k$, i.e. $n_k$. One sees that in the limit $k \to \infty$, the quantity $k^4 n_k$ converges to $4\pi m {\cal C}/\hbar V = C$ as consistently given by (\ref{C}) and showing, within  this model, the equality predicted by Tan, see eq.(\ref{c1}). Because of this identity, measurement of the contact\cite{Partridge-PRL2005,Stewart-PRL2010} may lead to an alternative determination of properties of the excitations energy spectra of these systems.

Before studying some properties of the contact, from the perspective of eq.(\ref{C}), we would like to point out that if we returned to the many-body description of the interacting fermion gas, in terms of the creation and annihilation operators of particles in states $k$, and then implemented the contact approximation of the interatomic potential, eq. (\ref{a1}), one would find that the sums are plagued by ultraviolet divergences. As Tan has suggested, a way to regularize the theory would be to introduce a kind of counterterm in the form $(\hat n_k - C/k^4)$. We find that in the mean-field BEC-BCS theory one can also regularize the theory at that level, namely, before implementing the variational scheme, and the counterterm is precisely Tan's prescription. At least at the level of a mean-field theory, we consider that this closes the circle of identifying the contact variable ${\cal C}$ as the thermodynamic conjugate of the inverse of the scattering length $\eta = 1/a$.\\

\subsection{Thermodynamics of the homogenous gas}

\begin{figure}
  \begin{center} \scalebox{0.8}
   {\includegraphics[width=\columnwidth,keepaspectratio]{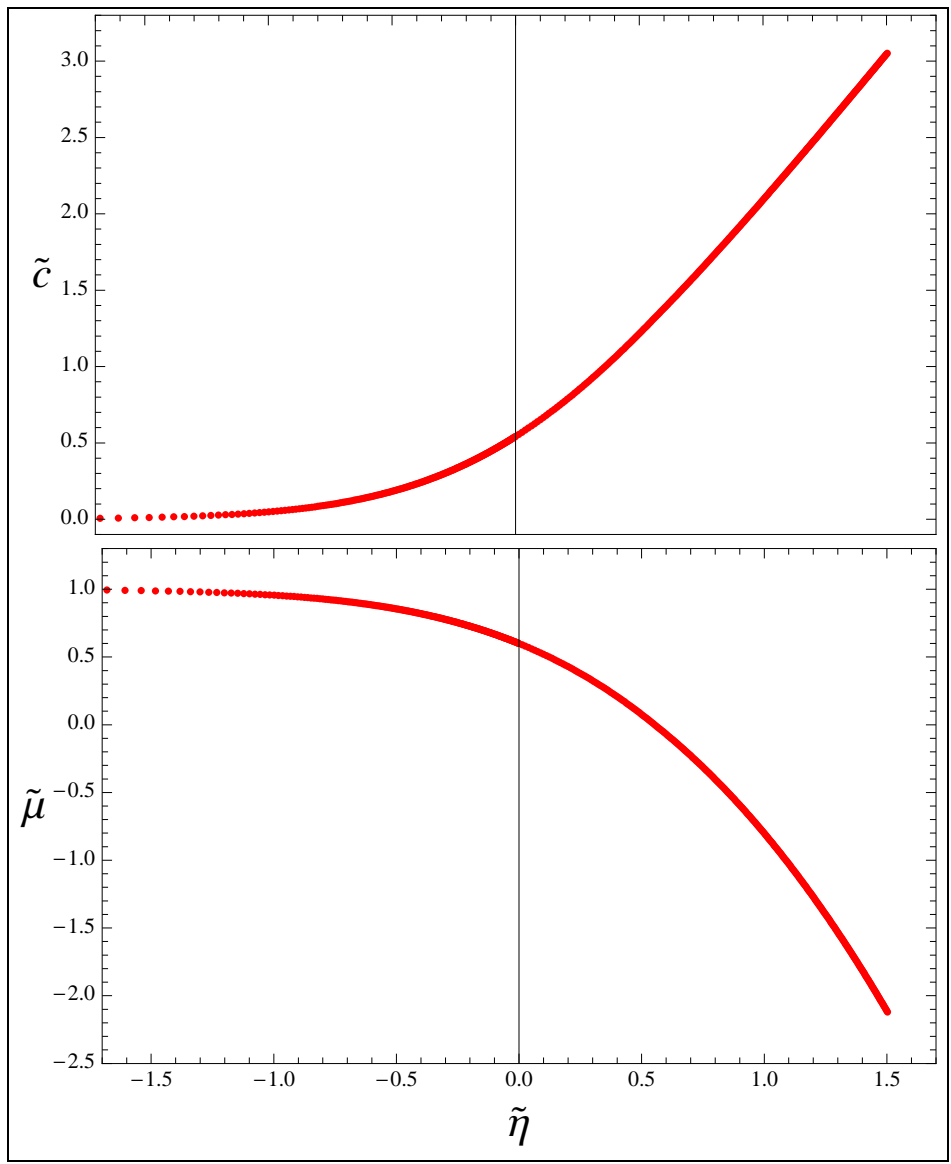}}
   \caption{Reduced contact $\tilde c$ vs $\tilde \eta$ and chemical potential $\tilde \mu$ vs $\tilde \eta$, for a homogenous fluid. $\tilde \eta = 1/k_F a$.}\label{fig1}
  \end{center}
\end{figure}

We first rewrite the three main equations, (\ref{gap}), (\ref{N}) y (\ref{C}), in terms of reduced variables, see eqs. (\ref{redh1}) and (\ref{redh2}),
\begin{equation}
\tilde \eta = - \frac{1}{\pi}  \int_0^\infty x^{1/2} d x \left[\frac{1}{\sqrt{(x - \tilde \mu)^2 + {\tilde \Delta}^2}} - \frac{1}{x}\right] \label{gap4}
\end{equation}
\begin{equation}
1 =\frac{3}{4}\int_0^\infty x^{1/2} d x 
\left[ 1 - \frac{x - \tilde \mu}{ \sqrt{(x - \tilde \mu)^2 + {\tilde \Delta}^2} } \right] \label{N4}.
\end{equation}
and
\begin{equation}
\tilde c = \frac{3 \pi }{8} {\tilde \Delta}^2 .\label{Cred}
\end{equation}
The solution to equations (\ref{gap4}) and (\ref{N4}) yield $\tilde \mu = \tilde \mu(\tilde \eta)$ and $\tilde \Delta = \tilde \Delta(\tilde \eta)$, as expected, and therefore, $\tilde c = \tilde c(\eta)$ as well. Fig. \ref{fig1} show $\tilde \mu$ vs $\tilde \eta$ and $\tilde c$ vs $\tilde \eta$. At unitarity $\tilde \eta = 0$ one finds $\tilde \mu(0) \approx 0.59$ and $\tilde c(0) \approx 0.54$. Typically, the value of the chemical potential at unitarity is related to the universal $\beta$ parameter\cite{Hu-NaP2007} as $\tilde \mu(0) = 1 + \beta$. From experiments\cite{Bourdel-PRL2004,Bartenstein-PRL2004,Partridge-Sc2006,Luo-PRL2007,Schirotzek-PRL2008} and quantum Monte Carlo simulations\cite{Carlson-PRL2003,Chang-PRA2004,Perali-PRL2004,Astrakharchick-PRL2004,Jauregui-JPB2010} as well as field theoretical calculations\cite{Hu-EL2006,Haussmann-PRA2007}, the accepted values are $\beta \approx - 0.64$ to $-0.49$ and $\tilde \Delta \approx 0.44$ to $0.54$. The mean field values are $\beta \approx - 0.41$ and $\tilde \Delta \approx 0.68$. It is well known that mean field is a bit off at unitarity, but as we see, such a simple theory allows for a full determination of the whole crossover yielding even semi-quantitative agreement with experimental data. 

One can also work out the asymptotic results at BCS, $\tilde \eta \to - \infty$, and at molecular BEC, $\tilde \eta \to + \infty$. These are, BCS,
\begin{equation}
\tilde \mu \approx 1 \>,\>\>\>\>\> \tilde c \approx 24 \pi e^{-4} \> e^{\pi \tilde \eta} \>\>\>\>\>{\rm as}\>\>\>\>\>{\tilde \eta \to - \infty},\label{BCSasi}
\end{equation}
and molecular BEC,
\begin{equation}
\tilde \mu \approx - \tilde \eta^2 + \frac{2}{3 \pi}\frac{1}{\tilde \eta} \>,\>\>\>\>\> \tilde c \approx 2 \tilde \eta \>\>\>\>\>{\rm as}\>\>\>\>\>{\tilde \eta \to + \infty} .\label{BECasi}
\end{equation}
Restoring the units of the first equation, it yields to lowest order at BEC, $\mu \approx - \hbar^2 /2m a^2$, half the binding energy of one molecule\cite{Leggett-80}. 

\begin{figure}
  \begin{center} \scalebox{0.8}
   {\includegraphics[width=\columnwidth,keepaspectratio]{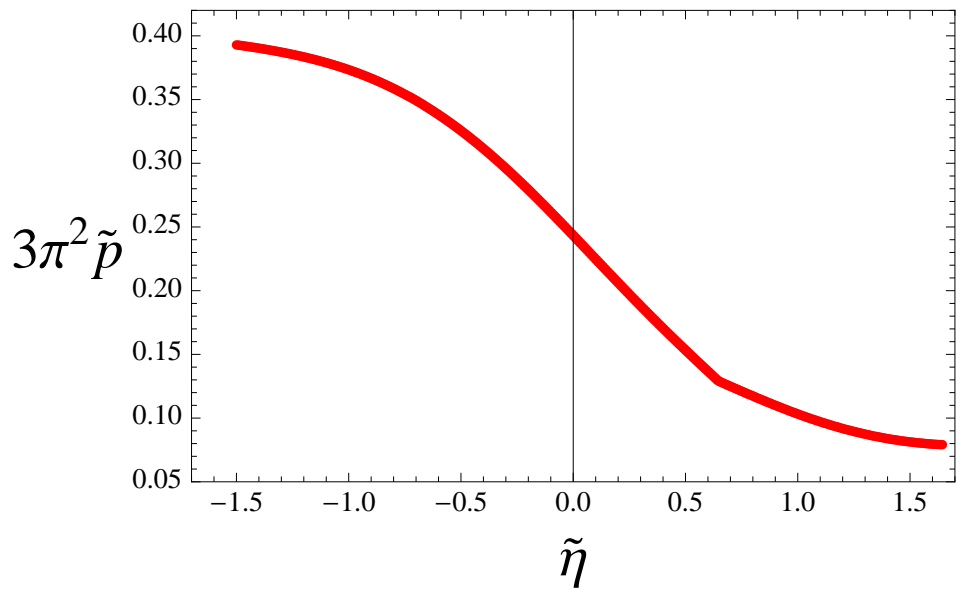}}
   \caption{Reduced pressure $\tilde p$ vs $\tilde \eta$ for a homogenous fluid. $\tilde \eta = 1/k_F a$. We have plotted $3 \pi^2 \tilde p$ to make direct comparison with the reduced pressure in the trap, see Fig.\ref{fig5}. See text for details.}\label{fig2}
  \end{center}
\end{figure}

To complete the thermodynamics we can calculate the pressure $p = -\Omega/V$. However, to make comparisons with the trapped case, we shall instead calculate the reduced grand potential $\tilde \omega = - 3 \pi^2 \tilde p$. This can be found directly from eq.(\ref{OMF}) in reduced units,
\begin{equation}
\tilde \omega = \frac{3}{4}  \int_0^\infty x^{1/2} d x \left[ (x - \tilde \mu) - \sqrt{(x - \tilde \mu)^2 + \frac{8}{3\pi}\tilde c} - \frac{4}{3\pi x}\tilde c \right] + \tilde c \> \tilde \eta .\label{o5}
\end{equation}
Since $\tilde \mu$ and $\tilde c$ are functions of $\tilde \eta$, so is $\tilde \omega$. The result is shown in Fig. \ref{fig2}. Although not shown here, the only remaining quantity is the internal energy. This is given by $\tilde e = \tilde \omega + \tilde \mu$.

\subsection{Thermodynamics of a  gas in a harmonic trap}

This case can be found resorting to LDA. First, we restore the units of the homogenous case to find the equation for the density profile in the trapped case, that is, we write $\mu = \mu(n,\eta)  = \epsilon_F \tilde \mu(\eta/k_F)$ and implement LDA: Take $\mu \to \mu - m\omega^2 r^2/2$ and $n \to \rho(\vec r)$, the latter being the density profile. One finds,
\begin{equation}
\mu - \frac{1}{2} m \omega^2 r^2 = \frac{\hbar^2}{2m} \left(3\pi^2 \rho(\vec r) \right)^{2/3} \> \tilde \mu\left(\frac{\eta}{(3 \pi^2 \rho(\vec r) )^{1/3}}\right) .\label{dplda}
\end{equation}
Then, the number of particles {\it in the trap} is found by integrating the density profile,
\begin{equation}
N = \int \> \rho(\vec r)\>d^3 r .\label{Nh}
\end{equation}
For the calculation of the contact, one must be careful to realize that one needs first the homogenous contact {\it density}, namely, $c_h={\cal C}/V$. This is,
\begin{equation}
c_h(n,\eta)  =  n \frac{\epsilon_F}{k_F} \tilde c\left(\frac{\eta}{(3 \pi^2 n)^{1/3}}\right) .
\end{equation}
Then one implements LDA to find the contact in the trap,
\begin{equation}
{\cal C} =  \int c_h(\rho(\vec r),\eta) \> d^3 r.
\end{equation}

\begin{figure}
  \begin{center} \scalebox{0.8}
   {\includegraphics[width=\columnwidth,keepaspectratio]{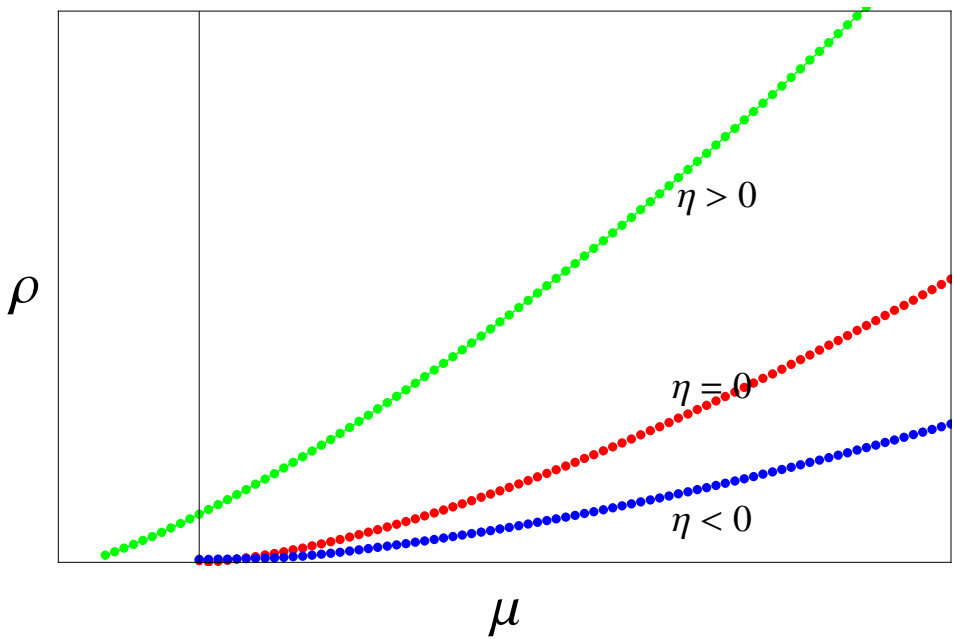}}
   \caption{Particle density $\rho$ vs chemical potential $\mu$, for a homogenous fluid and for three different values of the inverse scattering length $\eta$, in arbitrary units.}\label{fig3}
  \end{center}
\end{figure}

Solution to these three equations yield $\rho = \rho(\vec r; \mu,\eta, {\cal V})$,  $N = N({\cal V},\mu,\eta)$ and ${\cal C} = {\cal C}({\cal V},\mu,\eta)$. To actually solve those equations one assumes $N/{\cal V} = N\omega^3$ fixed and solves separately the cases $\eta \le 0$ and $\eta > 0$. This last separation is useful since one can show, using the form of $\tilde \mu$ as given in Fig. \ref{fig1}, that for fixed $\eta \le 0$, as $\mu \to 0^{+}$, $n \to 0$ and remains zero as $\mu$ becomes negative, while for fixed $\eta > 0$, $n \to 0$ as $\mu \to -{\hbar^2\eta^2}/{2m}$. This is shown in Fig. \ref{fig3}. This separation allows to determine the range of values of the density profiles in the trap. That is, one can see that in general $\rho(r) \to 0$ as $r \to r_{max}$, where $r_{max}$ is given by,
\begin{equation}
r_{max} = \left\{
\begin{array}{ccc}
\left(\frac{2\mu}{m\omega^2}\right)^{1/2} &{\rm if}& \eta \le 0 \\
\left( \frac{2 \mu}{m \omega^2} + \frac{\hbar^2 }{m^2 \omega^2}\eta^2 \right)^{1/2}&{\rm if}& \eta > 0 
\end{array} \right.  .\label{rmax}
\end{equation}

Using now reduced ``hat" variables for the trap, see eqs.(\ref{redt1}) and (\ref{redt2}), the equations to solve, (\ref{dplda}) and (\ref{Nh}), become, for the density profile,
\begin{equation}
\left\{ \begin{array}{c}
\hat \mu (1 -  x^2) \\ \hat \mu  - \left(\hat \mu + \hat \eta^2\right) x^2 
\end{array} \right\} =
  \left(3\pi^2 \hat \rho(x) \right)^{2/3} \> \tilde \mu\left(\frac{\hat \eta}{(3 \pi^2 \hat \rho(x) )^{1/3}}\right) ,\label{dplda3}
\end{equation}
for the total number of atoms,
\begin{equation}
1 = 96 \pi \> \left\{ \begin{array}{c}
\hat \mu^{3/2} \\ \left( \hat \mu + \hat \eta^2\right)^{3/2}
\end{array} \right\}  \> \int_0^{1} \> x^2 \> \hat \rho(x; \hat \mu,\hat \eta) \> dx ,
\label{NH3}
\end{equation}
and for the contact,
\begin{equation}
\hat c =
\frac{32}{\pi}\> \left\{ \begin{array}{c}
\hat \mu^{3/2} \\ \left( \hat \mu + \hat \eta^2\right)^{3/2}
\end{array} \right\} \> \int_0^1 \> x^2 dx\> \left(3 \pi^2 \hat  \rho(x) \right)^{4/3} \> \tilde{c} \left(\frac{\hat \eta}{(3 \pi^2 \hat \rho(x) )^{1/3}}\right) .\label{hatC}
\end{equation}
In the above three equations, the upper term in braces is for $\hat \eta \le 0$ and the lower one for $\hat \eta > 0$. Note that the profile becomes $\hat \rho = \hat \rho(x; \hat \eta)$, with $0\le x \le 1$. It is also clear that $\tilde \eta$ is the only independent variable, namely, $\hat \mu = \hat \mu (\hat \eta)$ and $\hat c = \hat c (\hat \eta)$. It is interesting to point out that the previous equations (\ref{dplda3}) - (\ref{hatC}) are valid in general, not only at mean-field level, since their derivation requires LDA and the boundary condition (\ref{rmax}) only. Mean-field enters as the model for $\tilde \mu(\tilde \eta)$ and $\tilde c(\tilde \eta)$.

\begin{figure}
  \begin{center} \scalebox{0.8}
   {\includegraphics[width=\columnwidth,keepaspectratio]{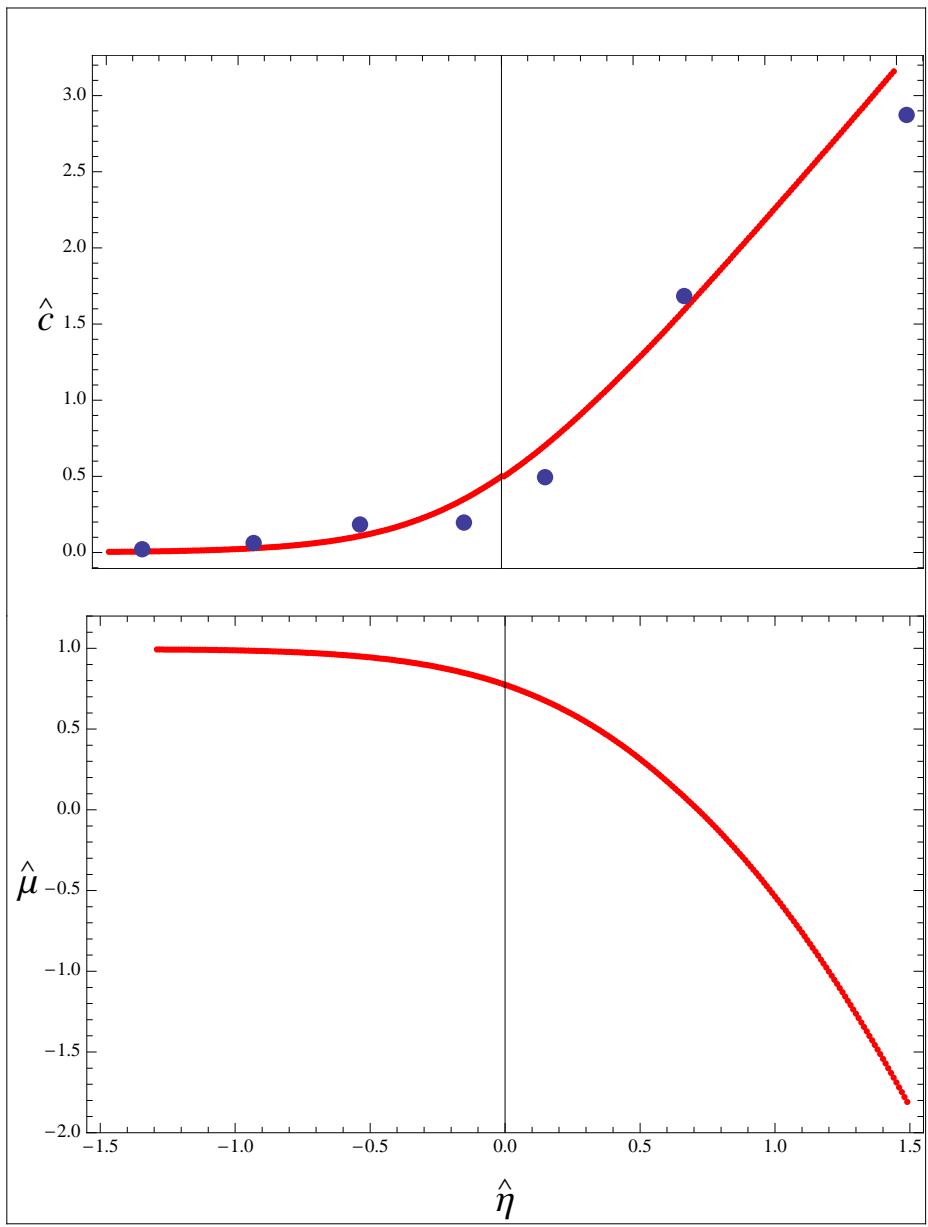}}
   \caption{Reduced contact $\hat c$ vs $\hat \eta$ and chemical potential $\hat \mu$ vs $\hat \eta$, for a fluid in a harmonic trap. $\hat \eta = 1/{\kappa}_F a$. The data points are from the experiments of Ref.\cite{Partridge-PRL2005}, as taken from Ref.\cite{Werner-EPJ2009}.}\label{fig4}
  \end{center}
\end{figure}

Equations (\ref{dplda3})-(\ref{hatC}) can be solved numerically, with the input from the homogeneous solution $\tilde \mu(\tilde \eta)$ and $\tilde c(\tilde \eta)$, and their solution $\hat \mu$ vs $\hat \eta$ and $\hat c$ vs $\hat \eta$ is shown in Figs. \ref{fig4}. One can find special and asymptotic behaviors. At unitarity $\hat \eta = \tilde \eta = 0$, and one finds 
\begin{equation}
\hat \mu(0) = (\hat \mu(0))^{1/2}, \label{hatmuuni}
\end{equation}
a known result, while the contact at unitarity,
\begin{equation}
\hat c(0) = \frac{256}{105 \pi} \frac{1}{\tilde \mu(0)^{1/4}} \tilde c(0)
.\label{hatCuni}
\end{equation}
a strange looking result, but quite interesting, since this relationship and (\ref{hatmuuni}) should be valid in general not only at the mean-field level. From the equations above one can check that these results follow by assuming universality at unitarity, LDA and the boundary condition $\rho(r) \to 0$ as $r \to r_{max} = (2\mu/m\omega^2)^{1/2}$. Mean-field provides the values $\tilde \mu(0)$ and $\tilde c(0)$ at unitarity, but these could be taken from experiments or more precise theoretical calculations. At mean-field level, we have shown that $\tilde c(0) \sim \tilde \Delta^2(0)$, and therefore, knowledge of the contact {\it in the trap} at unitarity yields information on the uniform gap at unitarity.

Given the approximated nature of mean-field, the calculated reduced contact in the trap $\hat c(\hat \eta)$ is in reasonable agreement with the more exact calculations by Werner et al.\cite{Werner-EPJ2009} and with the recent experiments by Partridge et al.\cite{Partridge-PRL2005} and Stewart et al.\cite{Stewart-PRL2010}. This is seen in Fig. \ref{fig4}, $\hat c$ vs $\hat \eta$, where we also plot experimental data from Ref.\cite{Partridge-PRL2005}, as taken from Ref.\cite{Werner-EPJ2009}, see their Fig. 1a. In Ref.\cite{Partridge-PRL2005} the number of bound pairs in the closed-channel was measured and in Ref.\cite{Werner-EPJ2009} this number was related to the contact; the experiment was performed in a gas of $^6$Li confined by a harmonic trap. On the other hand, in Ref. \cite{Stewart-PRL2010} a comparison of experimental contact data is made with the theory of Ref.\cite{Werner-EPJ2009}, showing good agreement as well. Thus, a similar agreement with mean-field results can be inferred by comparing them with the theoretical results of Fig. 1a of Ref.\cite{Werner-EPJ2009}. Certainly, at unitarity mean-field does not quite agree as expected, but the theory rings true as fas as the overall physics is concerned. 



The other interesting results are the asymptotic values in the inhomogeneous case. In the extreme molecular BEC these are,
\begin{equation}
\hat \mu \approx - \hat \eta^2 + \left(\frac{5}{32 \hat \eta}\right)^{5/2} \>,\>\>\>\>\> \hat c \approx 2 \hat \eta \>\>\>\>\>{\rm as}\>\>\>\>\>{\hat \eta \to + \infty} .\label{BEChat}
\end{equation}
Restoring the units for $\hat \mu$, it yields to lowest order, $\mu \approx - \hbar^2 /2m a^2$, again, half the binding energy of one molecule\cite{Leggett-80}. For the BCS side we find,
\begin{equation}
\hat \mu \approx 1 \>,\>\>\>\>\> ln \hat c \approx \pi \hat \eta \>\>\>\>\>{\rm as}\>\>\>\>\>{\hat \eta \to - \infty} ,\label{BCShat}
\end{equation}
 Comparing with the corresponding equations for the homogenous case, eq.(\ref{BCSasi}) and (\ref{BECasi}), we find that $\hat c$, $\tilde c$, $\hat \mu$ and $\tilde \mu$ have the same dependence on their corresponding reduced $\tilde \eta$ or $\hat \eta$ to leading order.  For the chemical potential, the equality between the homogenous and the trapped case is expected, namely, it should be true that $\mu \to \epsilon_F$ or $\mu \to \varepsilon_F$ in the BCS side, while $\mu \to - \hbar^2 /2m a^2$ in the BEC extreme, regardless of the confining potential. However, the situation is somewhat unexpected for the contact. That is, one finds that asymptotically at BEC and at BCS, the reduced contact is related to the reduced inverse scattering length in the same functional form for both cases. In the uniform gas, the contact is directly related to the gap, see eq.(\ref{C}), however, in the inhomogeneous trapped case the gap should be a local function and, therefore, the contact in the trapped case is an integrated square gap, namely,
\begin{equation}
{\cal C}({\cal V},\mu,\eta) = \frac{m}{4 \pi \hbar^2}\int \Delta^2(\vec r) \>d^3 r .
\end{equation}
Even though the gap is a local quantity for a gas in a trap, indirect measurements of it have been alluded to in Refs.\cite{Partridge-PRL2005,Stewart-Na2008,Schirotzek-PRL2008}, without fully considering its local nature, but certainly showing consistency with theory and their own data interpretation. Although this certainly needs a more careful analysis, we believe the reason behind this consistency may be both, the fact that one may ascribe a kind of ``mean" gap in a trap, and that the dependence of the contact with $\eta$ is very similar in the homogenous and the trapped cases.



\begin{figure}
  \begin{center} \scalebox{0.8}
   {\includegraphics[width=\columnwidth,keepaspectratio]{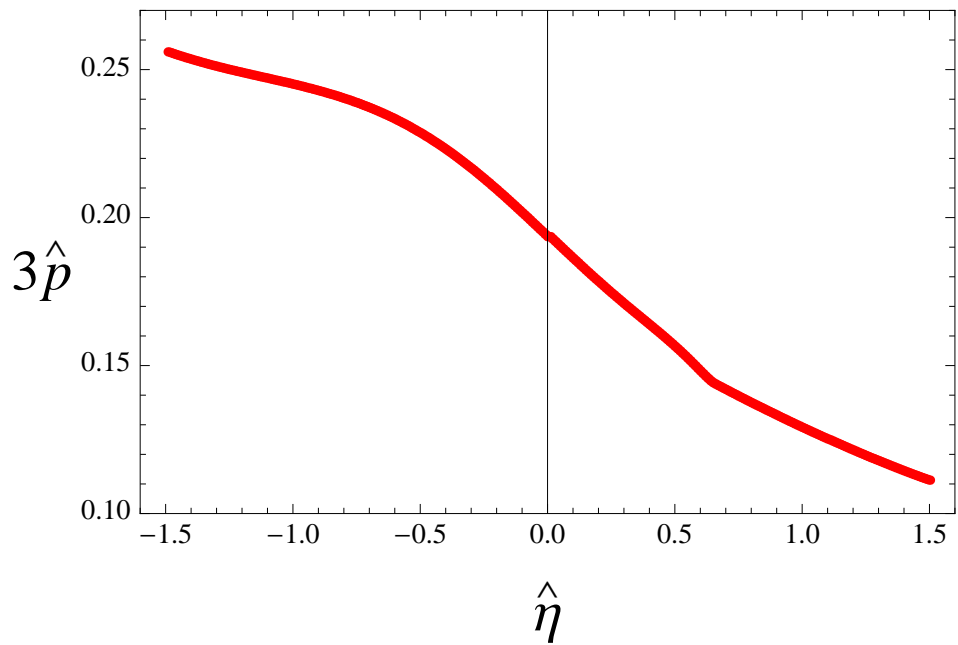}}
   \caption{Reduced generalized pressure $\hat p$ vs $\hat \eta$ for a homogenous fluid. $\hat \eta = 1/{\kappa}_F a$. We have plotted $3 \hat p$ to make direct comparison with the homogenous pressure, see Fig.\ref{fig3}. See text for details.}\label{fig5}
  \end{center}
\end{figure}

To end this section, and because this suggests an independent and useful experimental determination, we calculate the corresponding generalized pressure for the trapped gas. As discussed at large in Refs.\cite{VRR-PRL2005,VRR-BJP2005,Sandoval-PRE2008}, the grand potential in the trap is $\Omega = - {\cal P V}$ thus defining a generalized, pressure ${\cal P}$, conjugate  to the generalized volume ${\cal V}$. This generalized pressure is the {\it bona-fide} counterpart of the hydrostatic pressure $p$ of the homogenous gas. The equation of state of the trapped gas can be given in terms of ${\cal P} = {\cal P}(N/{\cal V},T,\eta)$ just as $p = p(N/V,T,\eta)$ in the homogenous one. It turns out that the measurement of ${\cal P}$ appears as a relatively simple task, given the enormous recent experimental advances. However, to make a direct comparison with the homogenous case, we will calculate the reduced grand potential $\hat \omega$ instead. One can show that $\hat \omega = - 3 \hat {\cal P}$. To perform this evaluation we could resort to integration of the homogenous grand potential density using LDA. However, as shown in  Refs.\cite{VRR-PRL2005,VRR-BJP2005,Sandoval-PRE2008}, one can instead use an extremely simple formula for the grand potential in the trap, namely,
\begin{equation}
\Omega = - \frac{2}{3} \int \rho(\vec r) \frac{1}{2} m \omega^2 r^2 \> d^3 r.
\end{equation}
That is, sole knowledge of the density profile and the external potential suffice to know the grand potential. We insist, if one has simultaneous knowledge of the number of particles $N$, the temperature $T$, and the scattering length $a$, then the equation of state of the system is immediately obtained. Using $T = 0$ and reduce units for the trap, we find  the following formula 
\begin{equation}
\hat \omega = - 64 \pi
\left\{ \begin{array}{c}
\hat \mu^{5/2} \\ \left( \hat \mu + \hat \eta^2 \right)^{5/2}
\end{array} \right\}
\int_0^1 x^4 \> \hat \rho(x; \hat \eta) \> dx ,
\end{equation}
where again the upper value is used for $\hat \eta \le 0$ and the lower one for $\hat \eta > 0$. The result is shown in Fig. \ref{fig5}. Comparing with Fig. \ref{fig2}, we find once more a remarkable similarity with the homogenous case.

\section{Final Remarks}
In this article we have performed a simple and straightforward derivation of the contact ${\cal C}$ as the thermodynamic conjugate variable to the inverse of the scattering length $a$. We have shown that essentially all the known results regarding the contact follow from this identification. We see that this identification is independent of any microscopic model one uses, as long as the scattering length can be varied externally. For instance, the state of the system may have any additional parametric dependence on other interatomic potential parameters and still the contact remains meaningful. Clearly, this is also valid for gases of Bose atoms. In the case where one can disregard other atomic parameters and keep the scattering length as the only two-body interaction dependence, dictated by the physical situation at hand, then a ``universal" or law of corresponding states emerges; namely, in the appropriate reduced variables, all thermodynamic variables depend on the reduced temperature and inverse scattering length.

Using the mean-field BEC-BCS theory, as introduced by Eagles\cite{Eagles-PR69} and Leggett\cite{Leggett-80}, one can explicitely exemplify all the above description. Furthermore, one finds a yet another identification of the contact, that is, its relationship to the (square) gap $\Delta$. This may serve as an additional and independent way of measuring or determining qualities of the atomic excitations spectra.  It is somewhat surprising that this approximated theory, in principle not valid in the crossover BEC-BCS region, still gives reasonable quantitative results, comparing well with both quantum Monte-Carlo calculations\cite{Carlson-PRL2003,Chang-PRA2004,Perali-PRL2004,Astrakharchick-PRL2004,Jauregui-JPB2010} as well as with actual experimental data\cite{Bourdel-PRL2004,Bartenstein-PRL2004,Partridge-Sc2006,Luo-PRL2007,Schirotzek-PRL2008,Stewart-PRL2010}. We would like to stress the fact that there is a remarkable similarity of the dependence of the reduced chemical potential, contact and pressure, on the reduced inverse of scattering length, for the homogenous and the trapped gases. This is not at all obvious from the outset and we wonder if this similarity is privative of the mean-field model or if it is a more general result. It is curious to note that, in this case of mean-field with sole dependence on the scattering length, the asymptotic values at the BEC and BCS extremes are more ``universal" than at unitarity! It remains as a useful excersice to work out all the thermodynamics here presented using field theoretical extensions of mean field, see Refs. \cite{Hu-EL2006} and \cite{Haussmann-PRA2007}, to obtain more precise values at unitarity.

In this article we have also pointed out the simplicity and, yet, the relevance of the measurement of the generalized pressure for trapped gases. Besides being readily measured just by knowing the density profile, the equation of state of the system is actually given in terms of this quantity, the frequency of the trap, the number of particles and the temperature. Its study and experimental determination may shed light on other properties, such as phase transitions and heat capacities\cite{Sandoval-PRE2008}. 

\ack{We acknowledge support from grant PAPIIT-UNAM IN116110.}

\end{document}